\documentstyle[12pt,fleqn]{article}
\begin{document}

\def\ai{\'{\i}}
\def\nu{n_{1}}
\def\nd{n_{2}}
\def\n3{n_{3}}
\def\Om{\Omega}

\newcommand{\be}{\begin{equation}}
\newcommand{\ee}{\end{equation}}
\newcommand{\ba}{\begin{eqnarray}}
\newcommand{\ea}{\end{eqnarray}}

\title{Integrability of the Pairing Hamiltonian}
\author{{M.C. Cambiaggio$^{1}$, A.M.F. Rivas$^{2}$\thanks{%
Corresponding author. e-mail:rivas@cbpfsu1.cat.cbpf.br, Tel:(5521)5410337
ramal195 fax:(5521)5412047 } and M. Saraceno$^{1}$}}
\maketitle

\noindent

\centerline{\it $^1$Departamento de F\'{\i}sica} \centerline{\it %
Comisi\'{o}n Nacional de Energ\'{\i}a At\'{o}mica (CNEA),} \centerline{\it %
Ave. del Libertador 8250, 1429 Buenos Aires, Argentina.} \centerline{\it $^2$%
Centro Brasileiro de Pesquisas F\'{\i}sicas} \centerline{\it Rua Xavier
Sigaud 150, CEP 22290-180, RJ, Rio de Janeiro, Brazil}

\vspace{1cm}


\centerline {\bf Abstract}

We show that a many-body Hamiltonian that corresponds to a system of
fermions interacting through a pairing force is an integrable problem , i.e.
it has as many constants of the motion as degrees of freedom. At the
classical level this implies that the Time-dependent Hartree-Fock-
Bogoliubov dynamics is integrable and at the quantum level that there are
conserved operators of two-body character which reduce to the number
operators when the pairing strength vanishes. We display these operators
explicitly and study in detail the three-level example.

\vfill
\clearpage

\section{Introduction}

Two kinds of simple models are commonly used in nuclear physics for
displaying the essential properties of the nuclear interaction, the
particle-hole and the particle-particle models. The simplest one, the Lipkin
model\cite{lip2}, consists of two levels of equal degeneracy and fermions
interacting through a particle-hole force. It has been extensively used to
test various approximation schemes and its classical version, provided by
the Time-Dependent Hartree -Fock (TDHF) approach, is integrable. When this
model is extended to three or more levels \cite{wilkoo,meredith,leboeuf} one
finds that the TDHF approximation yields a classical problem which is non
integrable, displays various degrees of chaotic behavior and has an
intricate and interesting phase-space structure \cite{leboeuf}.

In principle the analogous situation for the particle-particle interaction
can be thought to behave in a similar way. A two-level model with a pairing
force \cite{hogasen} is integrable \cite{camdus,broglia} and one would
expect that the extension to three or more levels would yield non-integrable
TDHF-Bogoliubov (TDHFB) dynamics. In this paper we report the fact that this
is not so and that the problem of a pairing force acting in a restricted
shell model space with $L$ single -particle levels turns out to be
integrable both classically and quantum mechanically. We display explicitly
the constants of the motion involved and we study their properties, their
group structure and their classical limits.

The outline of the paper is as follows. In Section 2 we review the pairing
model, in both its quantum and classical aspects. The classical limit is
obtained by a large scale degeneracy argument that leads to the TDHFB
equations of motion. Section 3 is devoted to the search for new constants of
the motion, i.e. new operators commuting with the Hamiltonian. These
constants of the motion, which are not unique, turn out to be non trivial
two body operators involving the coupling constant and the single particle
energies. A set of commuting operators is thus constructed that renders the
problem integrable. In Section 4 we treat the three- level case and show
explicitly the consequences of this integrability both at the quantum and
classical levels. The last Section is devoted to conclusions and final
remarks.

\section{The model}

The pairing force is a very general interacting mechanism that has an
ubiquitous role in the quantum many body problem. In electron systems it
leads to the superconducting mechanism and in nuclear physics to the
collectivity associated to the pairing degree of freedom. In this latter
case it provides a simplified description of the short range part of the
nuclear interaction \cite{borsom}. A schematic model that incorporates this
basic mechanism can be defined by interacting fermions that can occupy $L$
different single-particle shells of degeneracy $2\Omega_{i}$ and
single-particle energies $\epsilon_{i}$. The fermions interact via a
monopole pairing force. The Hamiltonian of such a system is 
\begin{equation}
H = \sum_{i=1}^{L} 2\epsilon_{i}K_{i}^{0}- \frac{G}{2}(K^{+} K^{-} + K^{-}
K^{+ }),  \label{eq:ham}
\end{equation}
where 
\begin{equation}
K_{i}^{0} = \frac{1}{2}\sum_{m_{i}}(b_{j_{i}m_{i}}^{\dagger}b_{j_{i}m_{i}} - 
\frac{1}{2}),
\end{equation}
\begin{equation}
K^+=\sum_{i=1}^L K_{i}^{+}=\sum_{i=1}^L \frac{1}{2} \sum_{m_{i}}
(-1)^{j_{i}- m_{i}} b_{j_{i}m_{i}}^{\dagger} b_{j_{i}-m_{i}}^{\dagger}
\end{equation}
and 
\begin{equation}
K^-=(K^+)^{\dagger}=\sum_{i=1}^L K_{i}^{-}.
\end{equation}
The $(b_{j_{i}m_{i}}^{\dagger},b_{j_{i}m_{i}}) $ are the usual fermion
operators obeying anticommutation relations which create or annihilate a
fermion on the i-th shell which has degeneracy $2j_{i}+1 = 2\Omega_{i}$. The 
$K_{i}^{0} $ operators count pairs of particles $N_{i}$ in each shell by $%
K_{i}^{0} = (N_{i} -\frac{1}{2} \Omega_{i} )$ . The operators $( K_{i}^{+}
,K_{i}^{-} ,K_{i}^{0} )$ conform an $SU(2) $ algebra whose Casimir operator
is 
\begin{equation}
k_{i}^{2} = K_{i}^{0^{2}} + \frac{1}{2} ( K_{i}^{+} K_{i}^{-} + K_{i}^{-}
K_{i}^{+}).
\end{equation}

%
%

The full dynamics of the system occurs in the group space of $\left[ SU(2)
\right]_{1} \times \left[ SU(2) \right] _{2} \times \ldots \times \left[
SU(2) \right] _{L} $.

The classical limit of the model is obtained from the TDHFB approximation
when the degeneracy $\Omega_{i}$ of each level goes to infinity \cite{camdus}%
. $\Omega_{i}^{-1}$ is the semiclassical parameter analogous to $\hbar$ in
the usual semiclassical treatments.

One way of obtaining this limit is through the time dependent variational
principle implemented through coherent states that are constructed from the
vacuum (or minimal weight) state $|0\rangle $, characterized by 
\begin{eqnarray}
K_i^{-}|0\rangle =0K_i^0|0\rangle =-\frac{\Omega _i}2|0\rangle
\end{eqnarray}
The coherent state in this representation is \cite{perelomov} 
\begin{equation}
|z\rangle =|z_1\ldots z_L\rangle =e^{\sum_{i=1}^L\bar{z}_iK_i^{+}}|0\rangle .
\end{equation}
The equations of motion obtained through the time-dependent variational
principle with this state are equivalent to the TDHFB equations. To obtain
them we use the variational principle appropriate for non-normalized states 
\cite{kramsar} with an action defined as (we set $\hbar =1$) 
\begin{equation}
S=\int dt\left[ \frac 12i\frac{\langle \psi |\dot{\psi}\rangle -\langle \dot{%
\psi}|\psi \rangle }{\langle \psi |\psi \rangle }-\frac{\langle \psi |H|\psi
\rangle }{\langle \psi |\psi \rangle }\right] .
\end{equation}
A detailed derivation is given in \cite{camdus}. The variables $z_i$ are not
canonical but the transformation 
\begin{eqnarray}
\omega _i=\sqrt{\frac{\Omega _i}{1+{z}_i\bar{z}_i}}{z}_i\mbox{   with  }%
0\leq \omega _i\bar{\omega _i}\leq \Omega _i
\end{eqnarray}
yields canonical variables satisfying $\{\omega _i,\bar{\omega _j}\}=\delta
_{ij}$. The finite range of these variables is a consequence of the Pauli
exclusion principle between correlated fermion pairs. In terms of $\omega _i$
the variational equations become ordinary hamiltonian equations in complex
form, 
\begin{eqnarray}
i\dot{\bar{\omega _i}}=\frac{\partial {\cal H}}{\partial \omega _i}%
\mbox{
and c.c },
\end{eqnarray}
where ${\cal {H}}=\lim_{\Omega _i\rightarrow \infty }\frac{\langle \psi
|H|\psi \rangle }{\langle \psi |\psi \rangle }$ is the classical hamiltonian
associated to the problem. In the same way for any quantum operator $\hat{A}$
its classical limit in this representation is, 
\begin{equation}
{\cal A}=\lim_{\Omega _i\rightarrow \infty }\frac{\langle \psi |\hat{A}|\psi
\rangle }{\langle \psi |\psi \rangle }.
\end{equation}

Therefore the operators $( K_{i}^{+} , K_{i}^{-}$ and $K_{i}^{0} )$ have
their classical analogues, $({\cal K}_{i}^{+} ,{\cal K}_{i}^{-} ,{\cal K}%
_{i}^{0} ) $, which can be written in terms of the $\omega_{i}$ variables 
\begin{eqnarray}
{\cal K}_{i}^{0} & = & \omega_{i}\bar{\omega_{i}} - \frac{\Omega_{i}}{2} \\
{\cal K}_{i}^{+} & = & \omega_{i}\sqrt{\Omega_{i} - \omega_{i}\bar{\omega_{i}%
}} \\
{\cal K}_{i}^{-} & = & \bar{\omega_{i}}\sqrt{\Omega_{i} - \omega_{i}\bar{%
\omega _{i}}}.
\end{eqnarray}
This last set of operators obey the classical $SU(2)$ Poisson bracket
relations \cite{kramsar}.

The classical pairing hamiltonian can then be expressed, in analogy with the
quantum one, directly in terms of the $SU(2)$ generators, 
\begin{equation}
{\cal H }= \sum_{i=1}^{L} 2\epsilon_{i}{\cal K}_{i}^{0}- G({\cal K^{+} K^{-}}
+ \sum_{i=1}^{L}\frac{\Omega_{i}}{2} ).  \label{eq:hcla}
\end{equation}
Energy conservation is guaranteed by the time-dependent variational
principle \cite{kerman} so that the motion occurs in the $(2L -1) $%
-dimensional manifold defined by ${\cal H}(\omega,\bar{\omega}) = E $. There
is a further constant of the motion linked to the conservation of the total
number of pairs \cite{camdus} 
\begin{equation}
{\cal N} (\omega,\bar{\omega}) = \sum_{i=1}^{L} \frac{<z|N_{i}|z>}{<z|z>} =
\sum_{i=1}^{L} \omega_{i}\bar{\omega_{i}}.  \label{eq:ni}
\end{equation}

The TDHFB equations are then classical hamiltonian equations in a phase
space of $2L$ variables. The conservation of ${\cal H }$ and ${\cal N }$
implies that the case $L=2$ is classically integrable, a fact that was
exploited in \cite{camdus} to compute energy levels and transition matrix
elements semiclassically.

\section{Search for additional constants of the motion}

The proof of the integrability for $L>2$ requires the existence of $L-1$
independent, well defined , global functions (constants of the motion),
whose Poisson brackets with each other and with the hamiltonian vanish.
Following Hietarinta\cite{hiet} a quantum mechanical hamiltonian of $L$
degrees of freedom is defined to be quantum integrable if there are $L-1$
independent, well defined, global operators (quantum invariants) which
commute with each other and with the hamiltonian. Then the energy spectrum
of a quantum integrable hamiltonian system is naturally labeled by $L$
quantum numbers \cite{niraj}, which are the eigenvalues of the corresponding
quantum invariants. Likewise the stationary states are simultaneous
eigenfunctions of the $L$ corresponding operators.

For the cases $L=1$ and $L=2$ the integrability is trivial. $H$ and $K_0$
(or $N$) provide the commuting operators and ${\cal H}$ and ${\cal K}_0$ (or 
${\cal N}$) the corresponding classical conserved quantities. But in the
case of $L>2$ no new obvious quantum invariants are present and we could
expect the system to be non integrable and therefore display generically
regions of chaotic behavior. However we now show how to construct non
trivial operators, independent of the hamiltonian and the total number of
pairs and commuting with them, which make the problem integrable.

For this purpose let us construct the more general two-body operator $O$
which is hermitian and conserves the total number of pairs 
\begin{equation}
R=\sum_{i,j}(\alpha _{ij}K_i^0K_j^0+\beta _{ij}K_i^{+}K_j^{-})+\sum_i\gamma
_iK_i^0
\end{equation}
with $\alpha _{ij}=\alpha _{ji}$ and $\beta _{ij}=\beta _{ji}$, and require
it to commute with the hamiltonian 
\begin{equation}
\left[ H,R\right] =0.
\end{equation}
The following relations among its coefficients 
\begin{eqnarray}
\alpha _{ij} &=&\beta _{ij} \\
\alpha _{ij} &=&-\frac G2\frac{\alpha _{ii}+\gamma _i-\alpha _{jj}-\gamma _j%
}{(\epsilon _i-\epsilon _j)},
\end{eqnarray}
are sufficient for the commutativity but they do not determine completely
the coefficients. Thus several solutions can be found, and one should
further check that the operator constructed is independent of $H$ and $N$.

For example if we choose 
\begin{eqnarray}
\alpha _{ii}=0\mbox{and}\gamma _i=\epsilon _i
\end{eqnarray}
we then would have 
\begin{equation}
R=\frac H2-\frac G2\left[ (K^0)^2+\sum_{i=1}^Lk_i^2\right]
\end{equation}
where 
\begin{equation}
k_i^2=(K_i^0)^2+\frac 12(K_i^{+}K_i^{-}+K_i^{-}K_i^{+})
\end{equation}
is the Casimir of the $SU(2)$ algebra in the ith-level. In this case $R$ is
not useful because it is not independent of the conserved magnitudes that we
already know.

However another choice 
\begin{eqnarray}
\alpha_{ii} = 0 & \mbox{and} & \gamma_{i} = \delta_{li}
\end{eqnarray}
leads to a set of operators 
\begin{equation}
R_{i} = K_{i}^{0} - G\sum_{j\ne i}^{L} \frac{\vec{k_{i}}.\vec{k_{j}}}{(
\epsilon_{i} - \epsilon_{j} ) }  \label{eq:o2q}
\end{equation}
with $\vec{k_{i}}.\vec{k_{j}} = K_{i}^{0} K_{j}^{0} + \frac{1}{2} (
K_{i}^{+} K_{j}^{-} + K_{i}^{-} K_{j}^{+}) $.

In the non interacting limit ($G=0$) these operators become the natural set
of commuting operators $\{ K_{i}^{0};i=1, \ldots, L \}$. A straightforward
but tedious calculation shows that 
\begin{eqnarray}
\left[ R_{i} ,R_{j} \right] = 0 & &\mbox{ $i,j=1 \ldots L $}
\end{eqnarray}
for any value of $G$ and $\epsilon_{i}$. On the other hand, it is easy to
see that $H$ and $N$ can be written in terms of the $R_{i}$ as 
\begin{equation}
N = \sum_{i=1}^{L} R_{i}
\end{equation}
and 
\begin{equation}
H = \sum_{i=1}^{L} 2\epsilon_{i}R_{i} + G(\sum_{i=1}^{L} R_{i})^{2} -
G\sum_{i=1}^{L} k_{i}^{2}.
\end{equation}

Therefore we have constructed a set of $L$ commuting operators which also
commute with the pairing hamiltonian and are number conserving. They extend
to the fully interacting case the trivial properties of the number operators
of the $G=0$ system. Consideration of these operators then demonstrate the
integrability of the quantum problem. The simultaneous eigenvalue equations 
\begin{equation}
R_{i} | \psi(\lambda_{1}\ldots\lambda_{L}) \rangle = \lambda_{i} |
\psi(\lambda_{1}\ldots\lambda_{L}) \rangle
\end{equation}
gives the eigenvalues of the hamiltonian as 
\begin{equation}
H = \sum_{i=1}^{L} 2\epsilon_{i}\lambda_{i} + G(\sum_{i=1}^{L}
\lambda_{i})^{2} -\frac{G}{4}\sum_{i=1}^{L} (\Omega_{i}^{2} -1).
\end{equation}
%
%
%
%
It should be noticed however that the actual solution is by no means
simplified by this knowledge. The operators $R_{i} $ are two body operators
as complicated in principle as the hamiltonian itself and they cannot be
used (except by diagonalizing them) to separate the hamiltonian into
invariant subspaces. however, as we will see, the simple fact of their
existence has very drastic implications on the structure of both eigenvalues
and eigenfunctions.

In the classical limit the associated operators for the set $\{ R_{i} ;i=1
\ldots L \} $ written in terms of the classical operators $({\cal K}_{i}^{+}
,{\cal K}_{i}^{-} ,{\cal K}_{i}^{0} )$ are

\begin{equation}
{\cal R}_{i} ={\cal K}_{i}^{0} - G\sum_{j\ne i}^{L} \frac{\vec{{\cal K}_{i }}%
.\vec{{\cal K}_{j}}}{(\epsilon_{i} - \epsilon_{j} ) }  \label{eq:o2cl}
\end{equation}
where $\vec{{\cal K}_{i}}.\vec{{\cal K}_{j}}= {\cal K}_{i}^{0} {\cal K}%
_{j}^{0 } + \frac{1}{2} ( {\cal K}_{i}^{+} {\cal K}_{j}^{-} + {\cal K}%
_{i}^{-} {\cal K}_{j}^{+}) $.

Taking into account (\ref{eq:ham}), (\ref{eq:hcla}), (\ref{eq:o2q}) and (\ref
{eq:o2cl}) we can see that the classical operators ${\cal R}_{i}$,${\cal N } 
$ and ${\cal H }$ have the same structure (in terms of the $SU(2)$ algebra
generators ) as their quantum analogs, except for additive constants. It is
then easy to see that 
\begin{eqnarray}
\{ {\cal H},{\cal R}_{i} \} = 0 & \{ {\cal N},{\cal R}_{i} \} = 0 \\
\mbox{ and} & \{ {\cal R}_{i} ,{\cal R}_{j} \} = 0. \\
\end{eqnarray}
Analogously to the quantum case, the mean number of pairs ${\cal N}$ and the
energy ${\cal H }$ are 
\begin{equation}
{\cal N}({\cal R}_{1}\ldots {\cal R}_{L}) = \sum_{i=1}^{L} {\cal R}_{i}
\end{equation}
and 
\begin{equation}
{\cal H }({\cal R}_{1}\ldots {\cal R}_{L})= \sum_{i=1}^{L} 2\epsilon_{i}%
{\cal R}_{i} + G(\sum_{i=1}^{L} {\cal R}_{i})^{2} -\frac{G}{4}
\sum_{i=1}^{L} \Omega_{i}^{2}.
\end{equation}
The functions ${\cal R}_{i}(\omega,\bar{\omega})$ are surfaces on the $2L$
dimensional phase space. The trajectories lie in the intersection of these
surfaces, which are $L$ dimensional tori labeled by the constants ${\cal R}%
_{i}$. Chaotic motion therefore cannot occur in this system.

\section{Manifestations of integrability in the three-level case}

In this section we restrict our analysis to the three-level case ($L=3$) and
equal degeneracy $\Omega _i=\Omega $ and show the consequences of the
integrable behavior, both in the quantum and in the classical solutions.

%
%
As in the previous section we will start with the quantum treatment.
Although the problem has three degrees of freedom we have already seen that
the total number of pairs $N$ in the system is a conserved magnitude. We can
then use this fact to reduce explicitly the dimensionality to two degrees of
freedom and therefore we only need to display another quantum invariant
operator to have an integrable quantum problem. We then choose it as $O_2$
which is a linear combination of the $R_i$ operators defined in the previous
section, 
\begin{equation}
O_2=\sum_{i=1}^3\epsilon _i^2R_i.
\end{equation}
%
%
The quantum integrability shows up clearly when we display a grid of the
simultaneous eigenvalues of $H$ and $O_2$. This is done in Fig. 1 for the
case where the total number of pairs is $N=\Omega =14$, which corresponds to
a set of $120$ levels. We find that the eigenvalues lie on a regular grid
that includes all eigenvalues. The fact that this grid is not parallel and
equally spaced reflects the fact that $O_2$ and $H$ are not action variables
that quantize at integer spaced values. Of course a point transformation to
a set $S_1$ and $S_2$ exists if $O_2$ and $H$ are independent. But we do not
construct this variables explicitly. It is clear however that the grid we
obtain is a smooth deformation of a regular one. This proves that the two
operators ($O_2$ and $H$) are independent and that their simultaneous
eigenvalues form a set of good quantum numbers\cite{niraj}.

For the classical analysis we introduce the non-interacting action-angle
coordinates, 
\begin{eqnarray}
n_{i} & = & \omega_{i} \bar{\omega_{i}} \\
\phi_{i} & = & arg(\omega_{i} )
\end{eqnarray}
where $n_{i}$ is the mean number of pairs in the level $i$ and is now a
continuous classical variable. In these variables the classical hamiltonian
is, 
\begin{eqnarray}
{\cal H} & = 2\epsilon_{1} n_{1} + 2\epsilon_{3} n_{3} - \Omega
(\epsilon_{1} + \epsilon_{3} ) - & G \left\{ n_{1}(\Omega - n_{1} ) + n_{2}
( \Omega - n_{2} ) + \n3 ( \Omega - \n3 ) \right. \\
& & + 2\sqrt{ n_{1} n_{2} } \sqrt{(\Omega - n_{1} )( \Omega - n_{2} )} \cos(
\phi_{1} - \phi_{2} )  \nonumber \\
& & + 2\sqrt{ n_{1} \n3 } \sqrt{(\Omega - n_{1} )( \Omega - \n3 )} \cos(
\phi_{1} - \phi_{3} )  \nonumber \\
& & \left. + 2\sqrt{ n_{2} \n3 } \sqrt{(\Omega - n_{2} )( \Omega - \n3 )}
\cos( \phi_{2} -\phi_{3} ) \right\}  \nonumber
\end{eqnarray}
where we have taken $\epsilon_{2} = 0 $ as the energy reference.

The analysis is best performed by explicit elimination of the conserved
quantity ${\cal N} = \sum_{i} n_{i} $. We introduce the canonical
transformation,

\begin{eqnarray}
I_1=\frac{n_1}\Omega &&\hspace{2cm}\theta _1=\phi _1-\phi _3 \\
I_2=\frac{n_2}\Omega &&\hspace{2cm}\theta _2=\phi _2-\phi _3  \nonumber \\
I_3=\frac 1\Omega (n_1+n_2+\n3)=1 &&\hspace{2cm}\theta _3=\phi _3.  \nonumber
\end{eqnarray}
We have scaled the variables so that they are all in the same range and
adopted the mid-shell value, i.e. ${\cal N}(\omega ,\bar{\omega})=\Omega $.
The problem is reduced to a space of two degrees of freedom whose effective
hamiltonian is, 
\begin{eqnarray}
{\cal H}(I_1,I_2,\theta _1,\theta _2) &=&2\epsilon _1\Omega I_1+2\epsilon
_3\Omega (1-I_1-I_2)-\Omega (\epsilon _1+\epsilon _3) \\
&-&G\Omega ^2\left\{ I_1(1-I_1)+I_2(1-I_2)+(I_1+I_2)(1-I_1-I_2)\right. 
\nonumber \\
&&+2\sqrt{I_1I_2}\sqrt{(1-I_1)(1-I_2)}\cos (\theta _1-\theta _2)  \nonumber
\\
&&+2\sqrt{I_1(1-I_1-I_2)}\sqrt{(1-I_1)(I_1+I_2)}\cos (\theta _1)  \nonumber
\\
&&\left. +2\sqrt{I_2(1-I_1-I_2)}\sqrt{(1-I_2)(I_1+I_2)}\cos (\theta
_2)\right\} .  \nonumber
\end{eqnarray}
In the above coordinates the classical version of $O$ is 
\begin{eqnarray}
{\cal O}_2(I_1,I_2,\theta _1,\theta _2) &=&\epsilon _1^2\Omega I_1-\epsilon
_3^2\Omega (I_1+I_2)+\frac \Omega 2(\epsilon _3^2-\epsilon _1^2) \\
&-&G\Omega ^2\left\{ \epsilon _1I_1(\frac 12-I_1)+\epsilon _3(I_1+I_2-\frac
12)(1-I_1-I_2)\right.  \nonumber \\
&&+\epsilon _1\sqrt{I_1I_2}\sqrt{(1-I_1)(1-I_2)}\cos (\theta _1-\theta _2) 
\nonumber \\
&&+(\epsilon _1+\epsilon _3)\sqrt{I_1(1-I_1-I_2)}\sqrt{(1-I_1)(I_1+I_2)}\cos
(\theta _1)  \nonumber \\
&&\left. +\epsilon _3\sqrt{I_2(1-I_1-I_2)}\sqrt{(1-I_2)(I_1+I_2)}\cos
(\theta _2)\right\} .  \nonumber
\end{eqnarray}

It can be explicitly verified that $\{{\cal O}_2,{\cal H}\}=0$. We have also
tested in the numerical integration of the equations of motion that both $%
{\cal H}$ and ${\cal O}_2$ are constants; and use this fact to control the
numerical accuracy of the solutions. Integrability is also apparent in
Poincar\'{e} sections, as Fig. 2 shows. Only separatrices (no chaotic
layers) are observed for any value of the coupling constant. The salient
feature of this figure is the appearance of forbidden regions, quite a
common occurrence in spin systems\cite{leboeuf}.

\section{Conclusions}

Our analysis shows that the simple pairing force is integrable both at the
quantum and at the classical levels. The reason why this simple feature has
escaped attention can be found in the fact that the pairing force has been
used almost exclusively in the context of quantum many-body physics and for
energies close to the ground state. Thus there are no significant
consequences for ground state or RPA modes, which are in any case
integrable. However the structure of the highly excited states and
eigenfunctions should show this consequence. For example, the fluctuation
properties of the eigenstates will follow a Poisson rather than a GOE
statistics \cite{bohigas}, while eigenfuntions will show the traces of
conserved quantities and will not be ergodic.

There are very few models of interacting particles that are integrable\cite
{calogero} and the fact that this simple model belongs to this class comes
as a surprise. We have reported some consequences in the TDHFB dynamics and
on the spectrum. Other features should follow naturally, for example the
statistics of levels should be very different than the particle-hole case%
\cite{meredith}. Also the addition of a small perturbation to this
hamiltonian should follow the KAM\cite{kam} perturbation scheme and not a
fluctuating spectrum characteristic of the perturbation of chaotic systems.
From the technical side it is quite useful to have at one's disposal a model
which takes into account important features of the nucleon-nucleon
interaction and which {\it remains} integrable for arbitrary values of its
strength. Thus perturbation expansions can have a steadying point which need
not be the non interacting fermion system.

\bigskip
We acknowledge stimulating discussions with Prof. O. Bohigas. This work was
supported in part by CONICET Pid 3233/92. A.M.F.R. gratefully acknowledges
scholarships from Comisi\'{o}n Nacional de Energ\'{\i}a At\'{o}mica
(Argentina) and from CLAF-CNPq (Brasil).

\pagebreak
\clearpage

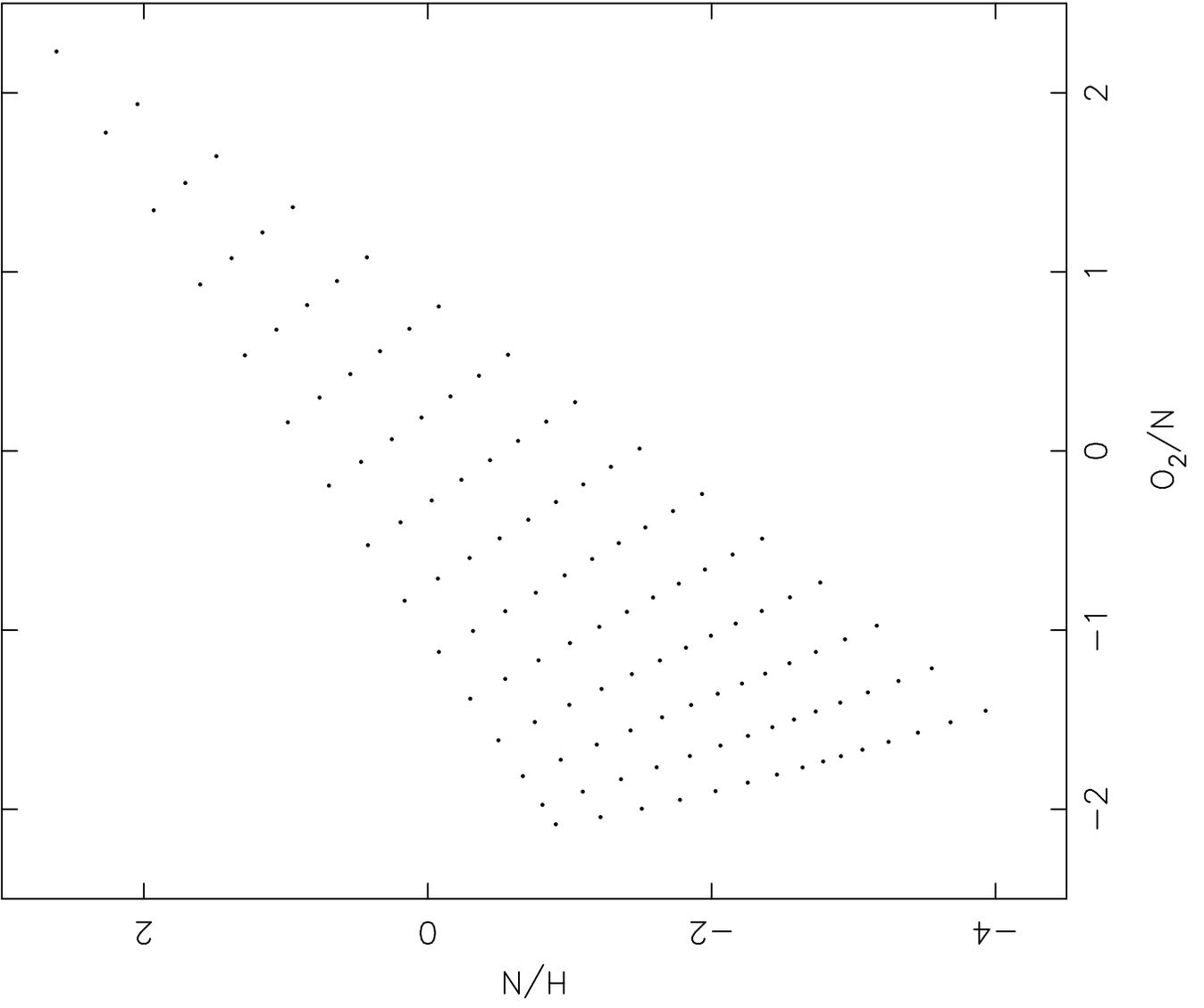
FIG. 1. Simultaneous eigenvalues of the Hamiltonian ($H$) and $O_2$ for $%
N=14 $ and $G=0.1$. The variables have been scaled dividing them by $N$.

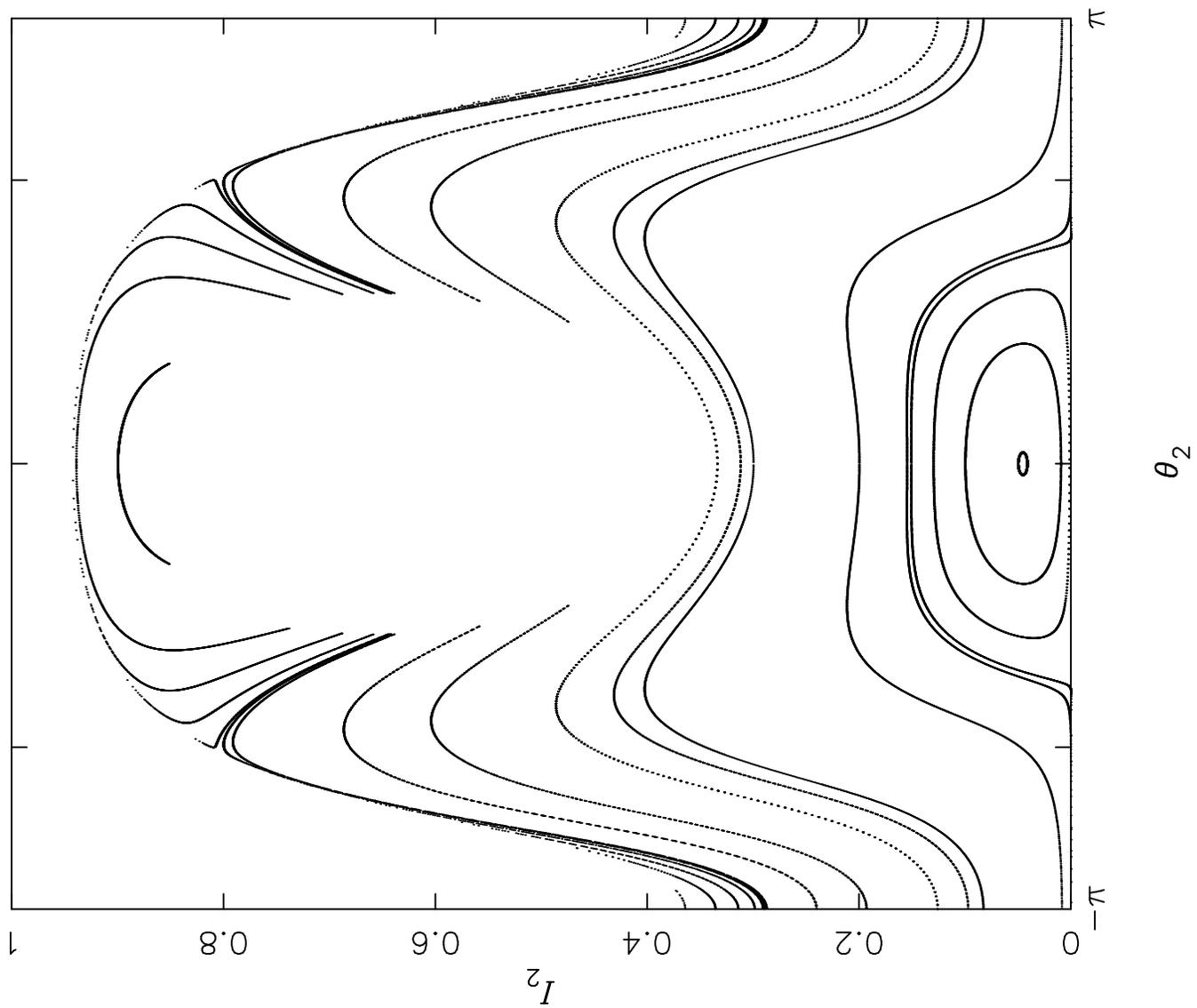
FIG. 2. Poincar\'e surface of section for the three-level pairing
hamiltonian . The mid-shell value ${\cal N} = \Omega $ has been taken and
the section is performed in the $I_{1} = 0$ case (see text).

\end{document}